
%
%
\input phyzzx

\rightline{hep-th/9507144, UTHEP-310}
\date{July, 1995}
\titlepage
\vskip 1cm
\title{\bf Prepotentials in N=2 SU(2) Supersymmetric Yang-Mills
Theory with Massless Hypermultiplets}
\author{Katsushi Ito and Sung-Kil Yang}
\address{Institute of Physics, University of Tsukuba,
Ibaraki 305, Japan}

\abstract{We calculate the prepotential of the low-energy effective action
for $N=2$ $SU(2)$ supersymmetric Yang-Mills theory with $N_f$ massless
hypermultiplets ($N_f=1,\, 2,\, 3$). The precise evaluation of the instanton
corrections is performed by making use of the Picard-Fuchs equations
associated with elliptic curves. The flavor dependence of the instanton
effect is determined explicitly both in the weak- and strong-coupling
regimes.}

\endpage
\overfullrule=0pt

\def\e{\hfill\break}
\def\half{{1 \over 2}}
\def\bra{\langle}
\def\ket{\rangle}
\def\L{\Lambda}

Low-energy effective action of $N=2$ supersymmetric Yang-Mills theory is
described in terms of a single holomorphic function
\REF\S{N. Seiberg, Phys. Lett. {\bf B206} (1988) 75}
[\S].
This type of
holomorphic function, called the prepotential, plays a vital role in
four-dimensional $N=2$ theories
\REF\Gates{S.J. Gates, Jr., Nucl. Phys. {\bf B238} (1984) 349; \e
B. De Wit and A. Van Proyen, Nucl. Phys. {\bf B245} (1984) 89}
[\Gates].
Seiberg and Witten discovered that
the prepotential for the gauge group $SU(2)$ is completely determined by
holomorphic data associated with elliptic curves
\REF\SWI{N. Seiberg and E. Witten, Nucl. Phys. {\bf B426} (1994) 19}
\REF\SWII{N. Seiberg and E. Witten, Nucl. Phys. {\bf B431} (1994) 484}
[\SWI,\SWII].
On the basis
of this seminal work, many non-perturbative aspects of the vacuum structure
of $N=2$ supersymmetric gauge theories have been revealed subsequently by
extending the gauge group
\REF\Many{A. Klemm, W. Lerche, S. Theisen and S. Yankielowicz,
Phys. Lett. {\bf B344} (1995) 169; \e
P. Argyres and A. Faraggi, Phys. Rev. Lett. {\bf 74} (1995) 3931; \e
M. Douglas and S. Shenker, {\it Dynamics of $SU(N)$ Supersymmetric Gauge
Theory}, hep-th/9503163; \e
U. Danielsson and B. Sundborg, {\it The moduli Space and Monodromies of
$N=2$ Supersymmetric $SO(2r+1)$ Theory}, hep-th/9504102; \e
P. Argyres and M. Douglas, {\it New Phenomena in $SU(3)$ Supersymmetric
Gauge Theory}, hep-th/9505062}
[\Many]
and by introducing matter hypermultiplets
\REF\Afew{A. Hanany and Y. Oz, {\it On the Quantum Moduli Space of $N=2$
Supersymmetric $SU(N_c)$ Gauge Theories}, hep-th/9505075;  \e
P. Argyres, M. Plesser and A. Shapere, {\it The Coulomb Phase of $N=2$
Supersymmetric QCD}, hep-th/9505100;  \e
J.A. Minahan and D. Nemeschansky, {\it Hyperelliptic Curves for
Supersymmetric Yang-Mills}, hep-th/9507032}
[\Afew].

In this paper we study the quantum moduli space of $N=2$
supersymmetric gauge theories with massless $N=2$ hypermultiplets of quarks.
The gauge group is $SU(2)$ and $N_f$ hypermultiplets are all in the spin
one-half representation of $SU(2)$. The theory is well-known to be
asymptotically free for $N_f \leq 3$. Our purpose is to calculate the
prepotentials of the low-energy effective actions for all the theories with
$N_f=1,\, 2$ and $3$.

Let $\phi$ be a complex scalar field
in the $N=2$ vector multiplet. The theory has a flat direction with
non-vanishing $\phi$, along which the gauge group $SU(2)$ is broken
to $U(1)$ and all the
quarks turn out to be massive. This is the Coulomb branch of the moduli
space. For $N_f \geq 2$ the other branch may develop when scalar fields in
the hypermultiplets acquire the vacuum expectation value. This branch is
called the Higgs branch where the gauge symmetry is completely broken.
In what follows we will concentrate on the Coulomb branch.

The quantum moduli space of the Coulomb branch is described by using the
gauge invariant order parameter $u= \bra {\rm Tr}\, \phi^2 \ket$.
The low-energy
effective theory at generic points in the complex $u$-plane contains an
$N=1$ $U(1)$ vector multiplet $W_\alpha$ and an $N=1$ chiral multiplet $A$
whose scalar component is denoted as $a$. The effective action is then
governed by a single holomorphic function ${\cal F}(A)$. We have
$$
{\cal L}={1 \over 4\pi} {\rm Im}
\Big[ \int d^4\theta {\partial {\cal F}(A) \over \partial A}\bar A
+\int d^2\theta {\partial^2 {\cal F}(A) \over \partial A^2}W_\alpha W^\alpha
\Big]
\eqno\eq
$$
in $N=1$ superspace. Let us define
$$
a_D={\partial {\cal F}(a) \over \partial a}.
\eqn\aDa
$$
Important insight in [\SWI] is to recognize the pair $(a_D, \, a)$
as a holomorphic section of an $SL(2,{\bf Z})$ bundle over the punctured
$u$-plane. The $SL(2,{\bf Z})$ acts on $(a_D, \, a)$ as the quantum
monodromy matrix around the singularities in the moduli space. At
singularities there appear extra massless states in addition to the
$N=2$ $U(1)$ vector multiplet.

It was found in [\SWI,\SWII] that the exact description of the moduli
space is determined through the elliptic curves. The curves are given by
$$
y^2=x^2(x-u)+{1 \over 4} \Lambda_0^4 \, x
\eqn\curvep
$$
for $N_f=0$ and
$$
y^2=x^2(x-u)-{1 \over 64} \Lambda_{N_f}^{2(4-N_f)} (x-u)^{N_f-1}
\eqn\curve
$$
for $N_f=1,\, 2,\, 3$. Here we have put all the bare quark masses to zero
and $\Lambda_{N_f}$ is the mass scale generated by the dimensional
transmutation.

In a recent paper
\REF\Klemm{A. Klemm, W. Lerche and S. Theisen, {\it Nonperturbative
Effective Actions of $N=2$ Supersymmetric Gauge Theories}, hep-th/9505150}
[\Klemm],
Klemm et al. calculated the prepotential for
$N_f=0$ explicitly. Their main tool is the Picard-Fuchs equations associated
with the $N_f=0$ elliptic curve which have also been considered in view
of special geometry
\REF\Cer{A. Ceresole, R. D'Auria and S. Ferrara, Phys. Lett. {\bf B339}
(1994) 71}
[\Cer].
For the present purpose we shall also employ the
technique of the Picard-Fuchs equation whose fundamental solutions are
combined so as to yield the exact expressions for $(a_D,\, a)$ in the
$N_f \geq 1$ theories.

In writing \curvep\ and \curve\
we have followed conventions in [\SWII] so that particles in the
hypermultiplets have integral electric charge. Hence the effective
coupling constant
$$
\tau(u) ={da_D \over da}
\eqno\eq
$$
is expressed as
$$
\tau(u) ={\theta_{eff}(u) \over \pi}+i{8\pi \over g^2_{eff}(u)}.
\eqno\eq
$$
The global symmetry acting on the $u$-plane is ${\bf Z}_2$ for $N_f=0$ and
${\bf Z}_{4-N_f}$ for $N_f \geq 1$. Notice that for $N_f=3$ there is no
symmetry in the $u$-plane. This fact distinguishes the $N_f=3$ theory from
the others.

Eq.\curve\ describes a double cover of the $x$-plane branched over
$e_0,\, e_{\pm}$ and $\infty$. We take a cut to run from $e_-$ to $e_+$
and from $e_0$ to $\infty$. Let us start with the $N_f=1$ curve. We have
for large $u$
$$
e_0 \simeq u, \hskip10mm e_\pm \simeq \pm i {\Lambda_1^3 \over 8 \sqrt{u}}.
\eqno\eq
$$
Thus the singularity at $u=\infty$ is not stable since
$e_0 \rightarrow \infty$
and $e_\pm \rightarrow 0$ as $u$ goes to infinity [\SWII].
Bearing this in mind
we make a rescaling $x \rightarrow ux$. We then obtain
$$
e_0={1 \over 3} (1+\xi+\xi^{-1}), \hskip10mm
e_\pm= {1 \over 3} (1 \mp \omega \xi \pm \omega^2/\xi),
\eqno\eq
$$
where $\omega =e^{i\pi/3}$ and
$$
\xi=2^{1/3}\big(2+27 \eta+3 \sqrt{3\eta(4+27\eta)}\big)^{-1/3}
\eqno\eq
$$
with $\eta=\Lambda_1^6/(64u^3)$. The singularities in
the $u$-plane are located
at $u=\infty$ and $u=-3 \Lambda_1^2 \omega^j/2^{8/3}$ with
$j=0,\, 1,\, 2$.

In the $N_f=2$ and $3$ curves the singularities are stable.
For $N_f=2$ we have
$$
e_0=u, \hskip10mm e_\pm =\pm \Lambda_2^2/8
\eqno\eq
$$
and the singularities in the $u$-plane are $u=\infty$ and $u=e_\pm$.
For $N_f=3$ we get
$$
e_0=u, \hskip10mm
e_\pm ={\Lambda_3^2 \over 128}
\Big(1\pm i\sqrt{{256 \over \Lambda_3^2}u-1}\Big)
\eqno\eq
$$
and hence the singularities occur at $u=\infty$, $u=0$ and
$u=\Lambda_{3}^{2}/256$.

The pair $(a_D,\, a)$ is now given by period integrals of the meromorphic
one-form $\lambda$,
$$
a_D=\oint_\beta \lambda, \hskip10mm a=\oint_\alpha \lambda.
\eqn\section
$$
Here the cycle $\alpha$ is taken to go around the cut connecting $e_-$ and
$e_+$ counter-clockwise.
The other cycle $\beta$ loops around the branch points
at $e_0$ and $e_-$ for $N_f=1$ and at $e_0$ and $e_+$ for $N_f=2,\,3$.
The intersection of $\alpha$ and $\beta$ is $\alpha \cdot \beta =1$.

The one-form $\lambda$ is determined by the differential equation [\SWII]
$$
{d\lambda \over du}={\sqrt{2} \over 8\pi} \lambda_1+d(*),
\eqn\difflam
$$
where $\lambda_1$ is the holomorphic differential $\lambda_1=dx/y$ and
$d(*)$ stands for the exact form in $x$. Integrating \difflam\ in $u$ we find
$$
\lambda ={\sqrt{2} \over 8\pi}\, {2u-(4-N_f)x \over y}\, dx
\eqn\oneform
$$
for $N_f\geq 0$. It is interesting to observe the one-loop beta function
coefficient $4-N_f$ in $\lambda$.

We wish to derive the Picard-Fuchs equation for the period
$\Pi=\oint \lambda$.
For this it is convenient first to write down the Picard-Fuchs equation for
the period $\Pi_1=\oint \lambda_1$. After some algebra we obtain
$$
p(u) {d^2 \Pi_1 \over du^2}+q(u){d\Pi_1 \over du}+\Pi_1  =0,
\eqn\pfholo
$$
where the coefficient functions are given by
$$
\eqalign{
N_f=1& \hskip10mm p(u)=4u^2+{27 \L_1^6\over 64u},
	\hskip10mm q(u)={1\over u^2}\Big(8u^3-{27 \L_1^6\over 64}\Big),\cr
N_f=2& \hskip10mm p(u)=4\Big(u^2-{\L_2^4\over 64}\Big),
	\hskip10mm q(u)=8u , \cr
N_f=3& \hskip10mm p(u)=u\Big(4u-{\L_3^2\over 64}\Big),
        \hskip10mm q(u)=8u-{\L_3^2\over 64}. \cr}
\eqn\coeff
$$
Combining \difflam\ and \pfholo\ it is immediate to see that
$$
{d^3 \Pi \over du^3}+{q(u)\over p(u)}{d^2\Pi \over du^2}
+{1 \over p(u)}{d\Pi \over du}  =0.
\eqn\pfthird
$$
Notice that the relation $q(u)=dp(u)/du$ holds for all $N_f$ in \coeff.
Consequently \pfthird\
reduce to the second order differential equations
$$
p(u){d^2 \Pi \over du^2}+ \Pi =0
\eqn\pf
$$
for $N_f\geq 1$. We have checked by verifying this equation directly
{}from \oneform\ without relying on \difflam.

Now that we have the Picard-Fuchs equations \pf, the periods \section\
will be expressed
by particular linear combinations of the two fundamental solutions to
\pf. In order to fix the combinations we have to evaluate the contour
integrals \section\ explicitly.
Let us first discuss the behavior at $u=\infty$.
For $N_f \geq 1$ the asymptotic freedom and the instanton contributions
lead one to expect that [\SWII]
$$
\eqalign{
a_D(u) &=i {4-N_f \over 2\pi} a(u) \ln {u\over \L_{N_f}^2}
+\sqrt{u}\sum_{n=0}^\infty
a_{Dn}(N_f) \Big({\L_{N_f}^2\over u} \Big)^{n(4-N_f)},  \cr
a(u) &=\half \sqrt{2u} \Big[1+\sum_{n=1}^\infty
a_n(N_f)\Big({\L_{N_f}^2\over u}\Big)^{n(4-N_f)} \Big] . \cr}
\eqn\SWformula
$$
Here the instantons with even instanton number $2n$ contribute the
terms $(\L_{N_f}^2/ u)^{n(4-N_f)}$. The amplitudes for the odd instanton
contributions vanish because of the anomalous ${\bf Z}_2$ symmetry for
$N_f \geq 1$ [\SWII]. From \SWformula\
 one can easily read off the monodromy matrix at $u=\infty$,
$$
M_{\infty}=\pmatrix{-1 & 4-N_f \cr
	   0  & -1    \cr}.
\eqno\eq
$$

We compute the lower order expansion of the integrals \section\ explicitly.
The results for $a_D(u)$ read
$$
a_D(u)=i {4-N_f \over 2\pi} \half \sqrt{2u}
\Big(\ln {u\over \L_{N_f}^2}+c_{N_f}\Big)+\cdots ,
\eqno\eq
$$
where $c_1=-{1\over 3}\ln e^{i\pi}+4\ln 2-2$, $c_2=6\ln 2-2$ and
$c_3=-\ln e^{i\pi}+12\ln 2-2$. For $a(u)$ we get
$$
a(u)=\half \sqrt{2u}\Big[1+a_1(N_f) \Big({\L_{N_f}^2\over u} \Big)^{4-N_f}
+\cdots \Big],
\eqno\eq
$$
where $a_1(1)=3/2^{10}$ and $a_1(2)=a_1(3)=-1/2^{10}$. Thus our explicit
results are in agreement with \SWformula.

The exact expressions for $(a_D,\, a)$ near $u=\infty$
are now obtained from the
solutions to the Picard-Fuchs equations.
The solutions to the Picard-Fuchs equations are written succinctly by
introducing a new variable
$$
z=c(N_{f}) \left( {u \over \Lambda_{N_{f}}^{2}}\right)^{4-N_{f}},
\quad
c(N_{f})=\eta_{N_{f}} 2^{8} (4-N_{f})^{N_{f}-4},
\eqn\va
$$
which keeps the discrete ${\bf Z}_{4-N_{f}}$ symmetry in the $u$-plane.
Here $\eta_{1}=-1,\, \eta_{2}=\eta_{3}=1$ .
The Picard-Fuchs equations \pf\ for $N_{f}\geq1$ become
the hypergeometric system with singularities at $z=0,1,\infty$
$$
z(1-z) {d^{2} \Pi \over d z^{2}}
+(\gamma-(\alpha+\beta+1)z){d \Pi \over d z}-\alpha\beta \Pi=0,
\eqn\hyperg
$$
where
$$
\alpha=\beta={-1\over 2(4-N_{f})}, \hskip10mm  \gamma={3-N_{f} \over 4-N_{f}}.
\eqno\eq
$$

The set of fundamental solutions to \hyperg\ near $u=\infty$ are
written as $w_{1}$, $w_{2}$
$$
\eqalign{
w_{1}(z)&=z^{{1\over 2(4-N_{f})}}
F\Big({1\over 2(4-N_{f})},{-1\over 2(4-N_{f})};1;{1\over z}\Big), \cr
w_{2}(z)&=w_{1}(z) \ln {1\over z} +z^{{1\over 2(4-N_{f})}}
F_{1}\Big({1\over 2(4-N_{f})},{-1\over 2(4-N_{f})};1;{1\over z}\Big),
\cr}\eqn\fund
$$
where
$$
\eqalign{
F(\alpha,\beta;\gamma;z)&=\sum_{n=0}^{\infty}
{(\alpha)_{n} (\beta)_{n}\over n! (\gamma)_{n}} z^{n}, \cr
F_{1}(\alpha,\beta;1;z)&=\sum_{n=0}^{\infty}
{(\alpha)_{n} (\beta)_{n}\over (n!)^{2}}
\sum_{r=0}^{n-1}\left({1\over \alpha+r}+{1\over \beta+r}-{2\over 1+r}
\right) z^{n}, \cr
\cr}\eqn\hype
$$
and $(\lambda)_{n}\equiv \lambda (\lambda+1) \cdots (\lambda+n-1)$.
Using this set of solutions, we are able to express $a(u)$ and $a_{D}(u)$
in the form:
$$
\eqalign{
a(u)&={ \Lambda_{N_{f}}\over \sqrt{2}c(N_{f})^{1\over 2(4-N_{f})}}
 w_{1}(z), \cr
a_{D}(u)&= -
{i\Lambda_{N_{f}}\over 2\sqrt{2}\pi c(N_{f})^{1\over 2(4-N_{f})}}
\big(w_{2}(z)+A_{N_{f}} w_{1}(z)\big),
\cr}\eqn\aad
$$
where $A_{N_{f}}=\ln c(N_{f}) -(4-N_{f}) c_{N_{f}}$.

In order to calculate the prepotential ${\cal F}$ we first express $u$ as
a series expansion in $a/\L_{N_f}$ by inverting $a(u)$. Substituting the
result into $a_D(u)$, and then integrating (2) with respect to $a$
we obtain the instanton expansion of ${\cal F}$,
$$
{\cal F}(a)={i a^{2}\over 2\pi}
\left\{ (4-N_{f}) \ln {a^{2}\over \Lambda_{N_{f}}^{2}}
+\sum_{k=0}^{\infty} {\cal F}_{k} (N_{f})
\left( {\Lambda_{N_{f}}^{2} \over a^{2}} \right)^{k(4-N_{f})}\right\}.
\eqn\inst
$$
The first five coefficients ${\cal F}_{k} (N_{f})$ are given by
$$
\eqalign{
{\cal F}_{1} (N_{f})&=-{1  \over 2^{b} c(N_{f})}
                         {1\over 2\, b^2}, \cr
{\cal F}_{2} (N_{f})&=-{1\over 2^{2b} c(N_{f})^{2}}
                       {5-8b+4b^{2} \over
                         2^{8}\, b^{4}}, \cr
{\cal F}_{3} (N_{f})&={1\over 2^{3b}c(N_{f})^{3}}
                      {   (1-b)(1-2b)
                         (-23+30b-16b^{2}) \over
                       1728\, b^{6}}, \cr
{\cal F}_{4} (N_{f})&={1\over 2^{4b} c(N_{f})^{4}}
{(1 - 2b )(-677 + 3910b - 8124b^2 +
      8456b^3 - 4672b^4 + 1152b^5) \over
294912\, b^8}, \cr
{\cal F}_{5} (N_{f})&=
(1 - 2b)(1 - b)
    (-7313 + 64386b - 210756b^2 + 335960b^3 \cr
 & -301792b^4 + 152704b^5 - 36864b^6))/
(2^{5b}c(N_{f})^{5} \  18432000 \, b^{10}),
\cr}\eqn\coef
$$
where $b=4-N_{f}$.
\REF\Matone{M. Matone, {\it Instantons and Reucrsion Relations in
$N=2$ SUSY Gauge Theory}, hep-th/9506102 }
One can obtain the $N_{f}=0$ result by replacing $c(N_{f})$ with 1
in the $N_{f}=2$ case, which recovers that in [\Klemm , \Matone].
Note that the coefficients ${\cal F}_{2n+1} (3)$ vanish for $n\geq 1$.

Our next task is to study the strong-coupling regime where the singularities
in the moduli space appear as a consequence of the extra massless monopole
(or dyon) states. Let us investigate the behavior near the singularity at
$u=u_0$ where the monopole becomes massless.
The position of the singularities is given
by $u_0=-3\L_1^2/2^{8/3}$ for $N_f=1$, $u_0=\L_2^2/8$ for $N_f=2$
and $u_0=0$ for $N_f=3$. In the $N_f \geq 1$ theory the monopole belongs
to the spinor representation of the global $SO(2N_f)$ symmetry [\SWII].
Hence, in the vicinity of $u=u_0$ , the effective dual $U(1)$ theory
consists of the $N=2$ $U(1)$ vector multiplet of magnetic photon and the
$k$ hypermultiplets of light monopoles. The multiplicity $k$ is given by the
dimension of the spinor
representation of $SO(2N_f)$, {\it i.e.} $k=2^{N_f-1}$.
Then the perturbation calculations in the dual theory predict [\SWII]
$$
\eqalign{
a & \simeq i{k \over 2\pi} a_D \ln a_D, \cr
a_D & \simeq c_0 (u-u_0), \hskip5mm c_0 \not= 0, \cr}
\eqn\dual
$$
{}from which we find the monodromy matrix at $u=u_0$,
$$
M_{u_{0}}=\pmatrix{1 & 0 \cr
	   -k & 1 \cr}.
\eqno\eq
$$

Let us turn to the explicit evaluation of the periods \section\ as $u$
approaches $u_0$. Our results for $a_D(u)$ read
$$
a_D(u)=c_D(N_f)\sqrt{2 v_{N_f}}\,
\Big[{u-u_0\over v_{N_f}}+b_D(N_f)\Big({u-u_0\over v_{N_f} }\Big)^2
+\cdots \Big],
\eqn\evalad
$$
where $v_{1}=-3\L_1^2/2^{8/3}$, $v_{2}=\L_2^2/4$,
$v_{3}=\L_3^2/64$, $c_D(1)=-i/4$, $c_D(2)=i/4$, $c_D(3)=1/4$, and
$b_D(1)=-1/24$, $b_D(2)=-1/8$, $b_D(3)=1/2$.

For $a(u)$ we obtain
$$
a(u)= { \sqrt{2 v_{N_f}} \over 2 \pi}
  \left\{ d(N_f)
+b(N_f)\, {u-u_0\over v_{N_f}}
\Big[\ln {u-u_0\over v_{N_f}} +d_{N_f} \Big] +\cdots \right\},
\eqn\evala
$$
where $d(1)=-3$, $d(2)=2$, $d(3)=i$, $b(1)=1/4$, $b(2)=-1/2$, $b(3)=i$
and $d_{1}=-4\ln 2-2\ln 3-1$, $d_{2}=-4\ln 2-1$, $d_{3}=-2\ln 2-1$. Notice
that
$$
{b(N_f) \over c_D(N_f)}=i2^{N_f-1}.
\eqno\eq
$$
Thus our results agree with the
perturbation result \dual\ with $k=2^{N_f-1}$ which is the right number of
massless monopole states.

Having checked the leading perturbative behavior we next discuss the exact
determination of $(a_D,\, a)$ with the use of the Picard-Fuchs equations.
In terms of the variable $z$ defined in \va\ the massless monopole points
are at $z=1$ for $N_f=1,\, 2$ and $z=0$ for $N_f=3$.
Define
$$
p(z)=\Bigg\{ \matrix{1-z, & \quad \ \ \  \hbox{for} \ \ N_{f}=1,\, 2 \cr
                            z, & \quad \hbox{for} \ \ N_{f}=3 \cr}
\eqno\eq
$$
The fundamental solutions to the Picard-Fuchs equation \hyperg\ near
$p(z)=0$ are then obtained as
$$
\eqalign{
w_{1}(z)&= p(z) F\Big(1-{1\over 2 b},
                          1-{1\over 2 b};2;p(z)\Big), \cr
w_{2}(z)&= w_{1}(z) \ln p(z) +p(z)
           F_{1}\Big(1-{1\over 2 b},
                          1-{1\over 2 b};2;p(z)\Big), \cr
}
\eqno\eq
$$
where $b=4-N_{f}$ and
$$
F_{1}(\alpha,\beta;2;z)={1\over (\alpha-1)(\beta-1)} {1\over z}
+\sum_{n=1}^{\infty}
{(\alpha)_{n} (\beta)_{n} \over n! (2)_{n}}
\sum_{r=0}^{n-1} \left({1\over \alpha+r}+{1\over \beta+r}-{1\over 1+r}
-{1\over 2+r}\right) z^{n}.
\eqno\eq
$$
In comparison with \evalad\ and \evala\ we find
$$
\eqalign{
a_{D}(u)&= C(N_{f}) w_{1}(z), \cr
a(u)    &= D(N_{f}) (w_{2}(z)+d_{N_{f}} w_{1}(z))
\cr}
\eqno\eq
$$
where
$$
\eqalign{
C(N_{f})&=-(-1)^{{N_{f}-1 \over 2}} (4-N_{f})^{-{2-N_{f}\over 2}}
     2^{-{5\over 2}-{3\over 2}(N_{f}-1)-{1\over 4-N_{f}}} \Lambda_{N_{f}}, \cr
D(N_{f})&={i\over 2\pi} 2^{N_{f}-1} C(N_{f}).
}
\eqn\coe$$

In the dual $U(1)$ theory the prepotential ${\cal F}_D$ is expressed as
a holomorphic function of $a_D$ and takes the form
$$
{\cal F}_{D}(a_{D})={i a_{D}^{2} \over 8\pi}
\left\{ k \ln \left( {a_{D}^{2} \over \Lambda_{N_{f}}^{2}} \right)
       +\sum_{n\geq -1} {\cal F}_{D \ n}(N_{f})
         \left( {a_{D} \over \Lambda_{N_{f}}} \right)^{n} \right\}.
\eqno\eq
$$
where $a=d {\cal F}_{D}/ d a_{D}$.
The expansion coefficients ${\cal F}_{D \ n}$ are obtained explicitly
in a similar way to the weak-coupling case.
After some computation we find
$$\eqalign{
{\cal F}_{D \ 1}(N_{f})&={2 \over 3 \tilde{c}(N_{f})}
                         {(1-2b)(-5+2b)\over 16 b^{2}}, \cr
{\cal F}_{D \ 2}(N_{f})&={1\over 2 \tilde{c}(N_{f})^{2}}
{(1-2b)(-61+190b-92 b^{2}+8b^{3})
\over 1152 b^{4}}, \cr
{\cal F}_{D \ 3}(N_{f})&={2\over 5 \tilde{c}(N_{f})^{3}}
{(1-2b)(1379-8162b+14596 b^{2}-7288b^{3}+960b^{4})
\over 110592 b^{6}}, \cr
}\eqn\instt$$
where
$
\tilde{c}(N_{f})=C(N_{f})/\Lambda_{N_{f}}.
$

Let us finally discuss other singularities
located at $u=u_*$ in the $u$-plane.
The $2\times 2$ quantum monodromy matrices $M_{u_*}$ have already been
obtained explicitly in [\SWII]. The left eigenvalue $(p,\, q)$ of $M_{u_*}$
gives the magnetic charge $p$ and the electric charge $q$
of the BPS dyonic state which
becomes massless at $u=u_*$. Except for the cace $u_*=u_0$ (massless monopole
point) we always have $q \not= 0$. Then, from the matrix $M_{u_*}$ one
can infer the behavior of $(a_D,\, a)$ near $u=u_*$, obtaining
$$
\eqalign{
a_D(u) & \simeq {1 \over 2\pi i} {m\over q}\
c_0 (u-u_*)\ln (u-u_*)+c_1 (u-u_*), \cr
a(u) & \simeq -{1 \over 2\pi i} {mp \over q^2}\ c_0 (u-u_*)\ln (u-u_*)+
{1\over q} (c_0-p c_1)(u-u_*), \cr}
\eqn\asymp
$$
where $c_0,\, c_1$ are non-vanishing constants and $m$ is the
$(1,2)$ component
of $M_{u_*}$. Notice that a good coordinate near $u=u_*$ is obtained as
$p  a_D+qa \simeq c_0(u-u_*)$.

We have checked the asymptotic formula \asymp\ for several cases of interest.
For instance, take a singularity at $u=\L^2_3/256 \equiv u_*$
in $N_f=3$ theory where the monodromy matrix is given by [\SWII]
$$
M_{u_{*}}=\pmatrix{3 & 1 \cr
           -4 & -1 \cr}
\eqno\eq
$$
and $(p,\, q)=(2,1)$. In order to find the desired expression \asymp\ one
has to follow carefully how the positions of the branch points (and hence the
branch cuts) in the $x$-plane evolve as $u$ approaches $u_*$ when evaluating
the contour integral \section. Letting $u\rightarrow u_*+0$ along the real
axis in the $x$-plane it is observed that the contour $\alpha$ is pushed to
the left of $u=u_*$ while $e_\pm$ approach $\L_3^2/128$ parallel to the
imaginary axis. In the limit $e_\pm \rightarrow \L_3^2/128$, therefore,
the evaluation of the period $a$ receives
the contribution of the $\alpha$-integral
(with the opposite sign) twice in addition to
the contour integral along the imaginary
axis. We thus arrive at
$$
\eqalign{
a_D & \simeq {1 \over 2\pi i}c_0 \epsilon \ln \epsilon +c_1\epsilon +c_2, \cr
a & \simeq -{1 \over 2\pi i} 2 c_0 \epsilon \ln \epsilon
+(c_0-2c_1)\epsilon -2c_2, \cr}
\eqno\eq
$$
where $\epsilon =1-256u/\L_3^2$ and $c_i$ are calculable constants.
This result
is in accordance with \asymp. Other cases are considered in a similar manner.

In summary, we have calculated the low-energy prepotenrial for $N=2$ $SU(2)$
supersymmetric Yang-Mills theory with massless hypermultiplets and determined
the flavor dependence of the instanton corrections exactly. It will be
interesting to study a similar issue when the bare quark masses are turned on.
In this case the elliptic curve takes the form
$$
y^{2}=x^3+ a (u) x^2 +b(u) x +c(u),
\eqn\ellip
$$
where $a(u), b(u), c(u)$ are functions of $u$ and the bare quark
masses [\SWII].
This leads to the set of Picard-Fuchs equations for the period integrals
of the first and second abelian differentials
$\Pi_{1}=\oint d x/y $ and $\Pi_{2}=\oint x d x/y $:
$$\eqalign{
{d \Pi_{1} \over d u}&= { p_{11}(u) \over  \Delta(u) }\Pi_{1}
                         +{p_{12}(u)  \over  \Delta(u)}\Pi_{2}, \cr
{d \Pi_{2} \over d u}&= { p_{21}(u) \over \Delta(u)} \Pi_{1}
                         +{  p_{22}(u)   \over  \Delta(u) }\Pi_{2}, \cr
}\eqn\mass$$
where
$$\eqalign{
p_{11}(u)&=-p_{22}(u)=a \,{{b }^2}\,a'  -
   4\,{{a }^2}\,c \,a'  +
   3\,b \,c \,a'  -
   2\,{{b }^2}\,b'  +
   6\,a \,c \,b'  +
   a \,b \,c'  - 9\,c \,c' ,  \cr
p_{12}(u)&= 2\,{{b }^2}\,a'  -
   6\,a \,c \,a'  -
   a \,b \,b'  + 9\,c \,b'  +
   2\,{{a }^2}\,c'  - 6\,b \,c' ,  \cr
p_{21}(u)&=-2\,{{b }^3}\,a'  +
   8\,a \,b \,c \,a'  -
   18\,{{c }^2}\,a'  +
   a \,{{b }^2}\,b'  -
   4\,{{a }^2}\,c \,b'  +
   3\,b \,c \,b'  -
   2\,{{b }^2}\,c'  +
   6\,a \,c \,c' , \cr
\Delta(u)&= 2\,\left( - {{a }^2}\,{{b }^2}  +
     4\,{{b }^3} + 4\,{{a }^3}\,c  -
     18\,a \,b \,c  + 27\,{{c }^2} \right) ,
\cr}
\eqno\eq$$
and the prime $'$ denotes the derivative with respect to $u$.
$\Delta(u)$ is the discriminant of the elliptic curve \ellip .
{} From \mass \ one can write down the third order differential equation
for $\oint \lambda$, which is analogous to \pfthird.
According to our preliminary analysis, however, the
equation will not reduce to the second order equation. Thus, one has to deal
with the monodromy property of the third order differential equation.

\vskip10mm

The work of K.I. is supported in part by University of Tsukuba
Research Projects and the Grant-in-Aid for Scientific Research from
the Ministry of Education (No. 07210210).
The work of S.-K.Y. is supported in part by Grant-in-Aid for Scientific
Research on Priority Area 231 ``Infinite Analysis'',
the Ministry of Education, Science and Culture, Japan.
\endpage

\refout

\end